**Thematic analysis of multiple sclerosis research by enhanced strategic diagram**

Nazlahshaniza Shafin[a], Che Aishah Nazariah Ismail[a], Mohd Zulkifli Mustafa[a], Nurhafizah Ghani[b], Asma Hayati Ahmad[a], Zahiruddin Othman[a], Adi Wijaya[c], Rahimah Zakaria[a*]

[a]School of Medical Sciences, Universiti Sains Malaysia, Kubang Kerian 16150, Malaysia
[b]School of Dental Sciences, Universiti Sains Malaysia, Kubang Kerian 16150, Malaysia
[c]Department of Health Information Management, STIKES Indonesia Maju, Jakarta, Indonesia

**\*Corresponding author:**

*Rahimah Zakaria, MBBS, PhD*

*Address: Department of Physiology, School of Medical Sciences, Universiti Sains Malaysia,*

*16150 Kubang Kerian, Malaysia,*

*Tel: +609 7676156,*

*E-mail: rahimah@usm.my*

**Running title**: Themes Analysis in Multiple Sclerosis Research

**Abstract**

This bibliometric review summarised the research trends and analysed research areas in multiple sclerosis (MS) over the last decade. The documents containing the term "multiple sclerosis" in the article title were retrieved from the Scopus database. We found a total of 18003 articles published in journals in the English language between 2012 and 2021. The emerging keywords identified utilising the enhanced strategic diagram were "covid-19", "teriflunomide", "clinical trial", "microglia", "b cells", "myelin", "brain", "white matter", "functional connectivity", "pain", "employment", "health-related quality of life", "meta-analysis" and "comorbidity". In conclusion, this study demonstrates the tremendous growth of MS literature worldwide, which is expected to grow more than double during the next decade especially in the identified emerging topics.

**Introduction**

Multiple sclerosis (MS) is an autoimmune chronic inflammatory, demyelinating, and neurodegenerative disease of the central nervous system.[1] MS is characterised by inflammation, demyelination, astroglial growth (gliosis) and neurodegeneration.[2] Reversible neurological deficits occur in the early phases of MS, characterised by clinically isolated syndrome and relapsing-remitting MS (RRMS), whereas secondary progressive MS (SPMS) is hallmarked by progressive neurological deficits and clinical disability.[1] Less than 10% of MS patients have primary progressive MS (PPMS), which means the disease has progressed from the outset.[1,2]

Across developed and developing countries, MS prevalence has climbed since 2013.[3] MS affects 2.8 million individuals worldwide (35.9% of the population), with females being twice as affected than males.[3] Most patients develop RRMS between the ages of 20 and 35, while those with PPMS develop it around the age of 40.[1] The exact cause of MS is unclear, although a combination of genetic and environmental risk factors is associated with it. Familial MS occurs in roughly 13% of all MS phenotypes.[4] MS is polygenic, meaning it is caused by polymorphisms in numerous genes, each of which increases the chance of illness. The HLA class I and II gene polymorphisms are the main risk factors for MS.[5] Besides Epstein–Barr virus (EBV) infection during adolescence and early adulthood, other environmental risk factors include lack of sun exposure, low vitamin D levels, and obesity during adolescence.[5]

Recent interest in MS research warrants an analysis of literature review to assess its current state and emerging research fields. An earlier bibliometric analysis on MS was published in 2014.[6] The authors examined MS research using articles published between 2003 and 2012.

Recently, a bibliometric analysis was performed that focused on highly cited articles from the Science Citation Index.[7] Our bibliometric evaluation, however, used MS literature from the Scopus database to identify main themes and topics based on author's keyword co-occurrence.

**Methods**

Data involved in this study were retrieved and downloaded from the Scopus database on 3rd August 2021. The search term used was "multiple sclerosis" in the article title within the time frame of 2012 to 2021. A total of 26648 documents were identified and were filtered for source type as "journal", document type as "article", and language as "English". Finally, 18003 documents were retrieved and downloaded for further analysis (Figure 1). The growth rate in the decade 2012–2020 (percentage change from 2012 to 2020) is calculated as follows: (papers published in 2020 – papers published in 2012/ papers published in 2012) × 100.[6]

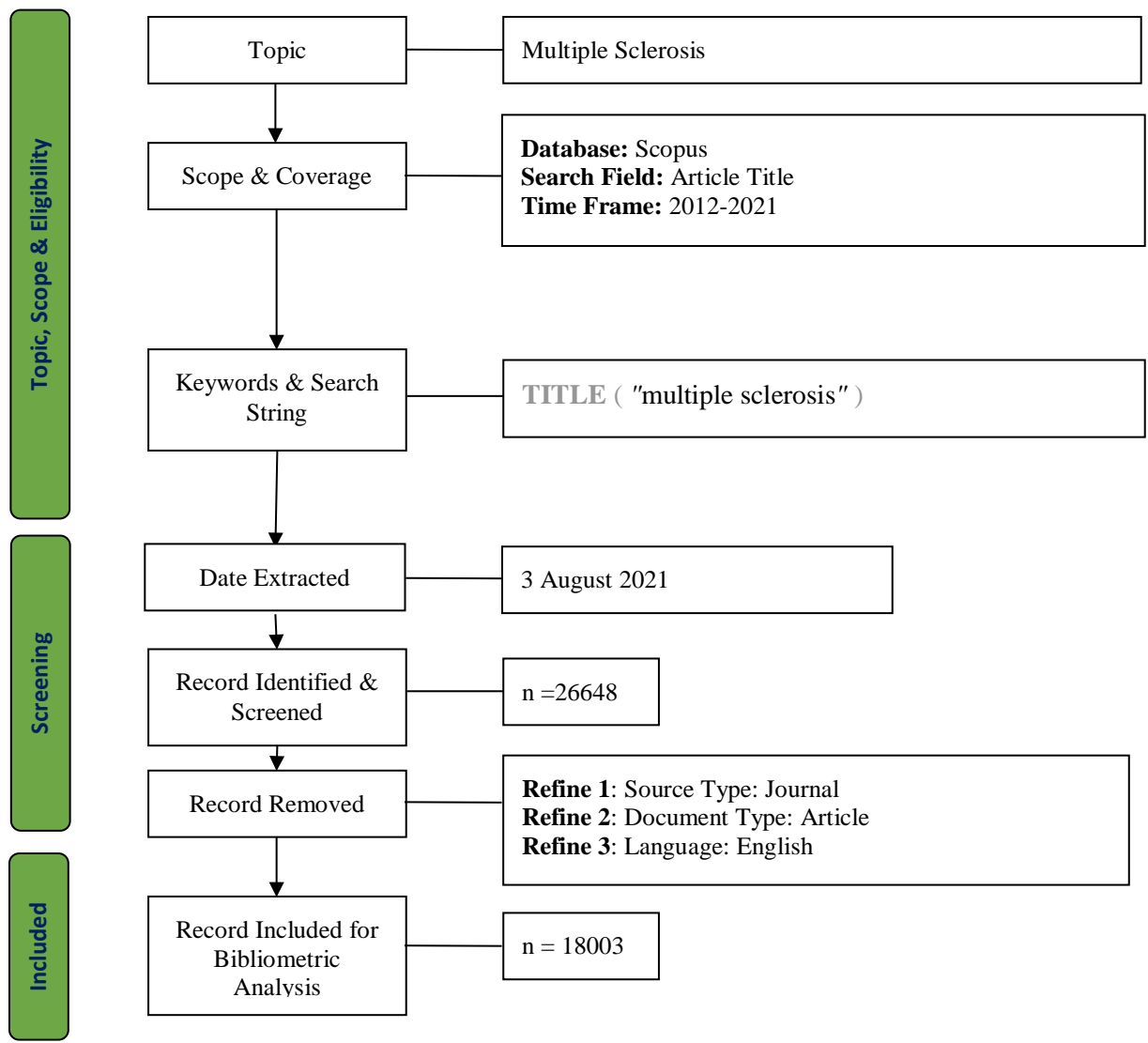

**Figure 1**. Search strategy used in this study.[8]

Harzing's Publish or Perish software was used to perform citation analysis, which used data in .ris format. It provided values for parameters such as TP=total number of publications; NCP=number of cited publications; TC=total citations; C/P=average citations per publication; C/CP=average citations per cited publication; h=h-index; and g=g-index.

VOSviewer software (1.6.17) was used to map the author's keywords, which used data in .csv format. It is a popular application with a simple graphic interface that can be used to build co-authorship network maps of authors, countries, citation analysis, and keywords co-occurrence.[9]

This method is especially beneficial for this study since its keyword co-occurrence analysis enables the identification of key research subjects and the detection of major research clusters linked to MS research. VOSviewer supports the introduction of a thesaurus file that can be used for data cleaning by integrating multiple forms of terms (e.g. biomarker and biomarkers, experimental autoimmune encephalitis and eae) to improve the correctness of the analysis.[10] The median values for occurrence, total link strength (TLS) and average publication year were calculated based on the data obtained from VOSviewer.

Theme networks, an overlay network and enhanced strategic diagrams (ESDs) were the three key results of our study. We have plotted the enhanced strategic diagram with the x-axis representing centrality, the y-axis representing density and the z-axis representing time, which is a three-dimensional plane based on a study by Feng et al.[11] The degree of interaction between networks is measured by centrality.[11,12] A theme with a higher centrality score has more external connections to other themes (external strength), thus has a greater impact on the field's development and evolution.[11-13] A node having 10 social connections, for example, would have a degree centrality of 10.[14] The higher the centrality score, the more central a topic is within the whole study field under consideration.[15] Thus, the mean strength value of external linkages to other subjects such as TLS was used in this study to determine centrality. The TLS values that were equal or higher than its median value were considered as high while those lower than its median values were considered as low centrality.

In contrast, the density of a topic is used to determine its internal strength or degree of interaction within a network.[11,12] In this study, co-occurrences of the author's keywords were used to determine the density. As for centrality, the median value was calculated and the co-occurrence values equal or higher than its median value were considered as high while those below it were considered as low density. The novelty of the topic, however, is represented by time[11] and in this study, the average publication year was used. In terms of novelty, the median value of average publication year was determined, and average publication years equal to or more than its median was deemed novelty, while those below it were considered old.

**RESULTS**

*The trend of annual publications from 2012 to 2021*

During the 2012–2021 period, 18003 articles on MS were retrieved from Scopus. In the first five years, 44.7% of these articles were published, while 55.3% were published in the second five years. There was a steady growth in the number of publications from 2012 through 2018 but rapid growth was observed in 2019. As to date, the highest number of articles was published in 2020 with a total of 7171 citations. Figure 2 shows the evolution of the number of publications and cited publications.

The overall growth rate of publications from 2012 to 2020 was 81.6%. Publications for 2021 was not considered in the calculation because of incomplete data. In terms of the number of cited publications, more than 90% of the articles published between 2012 and 2018 were cited. However, for articles published in 2019, 2020 and 2021, the percentage of cited publications was lower, at 87.8. 67.9 and 25.3, respectively.

**Table 1.** Annual number of publications and citation matrix.

| Year | TP | NCP | TC | C/P | C/CP | *h* | *g* |
|---|---|---|---|---|---|---|---|
| 2012 | 1359 | 1297 | 45276 | 33.32 | 34.91 | 89 | 143 |
| 2013 | 1574 | 1506 | 41785 | 26.55 | 27.75 | 84 | 121 |

| 2014 | 1649 | 1566 | 39145 | 23.74 | 25.00 | 76 | 111 |
| 2015 | 1702 | 1589 | 33812 | 19.87 | 21.28 | 69 | 106 |
| 2016 | 1770 | 1669 | 30845 | 17.43 | 18.48 | 63 | 97 |
| 2017 | 1831 | 1697 | 27709 | 15.13 | 16.33 | 53 | 94 |
| 2018 | 1897 | 1739 | 20476 | 10.79 | 11.77 | 47 | 71 |
| 2019 | 1978 | 1736 | 13952 | 7.05 | 8.04 | 36 | 54 |
| 2020 | 2468 | 1676 | 7171 | 2.91 | 4.28 | 24 | 37 |
| 2021 | 1775 | 449 | 930 | 0.52 | 2.07 | 8 | 12 |

Notes: TP=total number of publications; NCP=number of cited publications; TC=total citations; C/P=average citations per publication; C/CP=average citations per cited publication; h=h-index; and g=g-index.

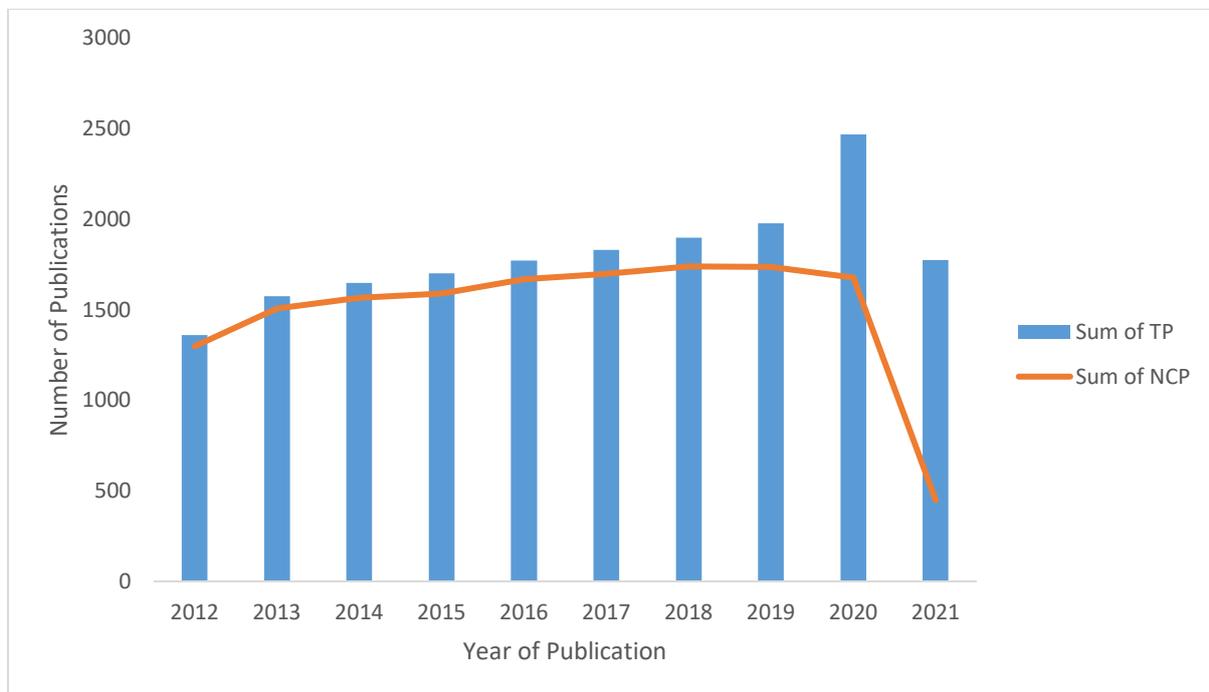

**Figure 2**. Evolution of the number of publications and cited publications by year.

*Topic analysis based on co-occurrence of author's keywords*

A co-word map was created with a threshold of 60 keyword co-occurrences in VOSviewer. To find themes in the literature, we studied each cluster's terms. Five themes emerge from the map in Figure 3: animal experiments, neuroimaging, psychosocial rehabilitation, epidemiology and management. The temporal co-word analysis identifies the publication dates when certain themes were at their most popular. Many newer keywords were identified under cluster 1 (red, management theme) compared to others (Figure 4).

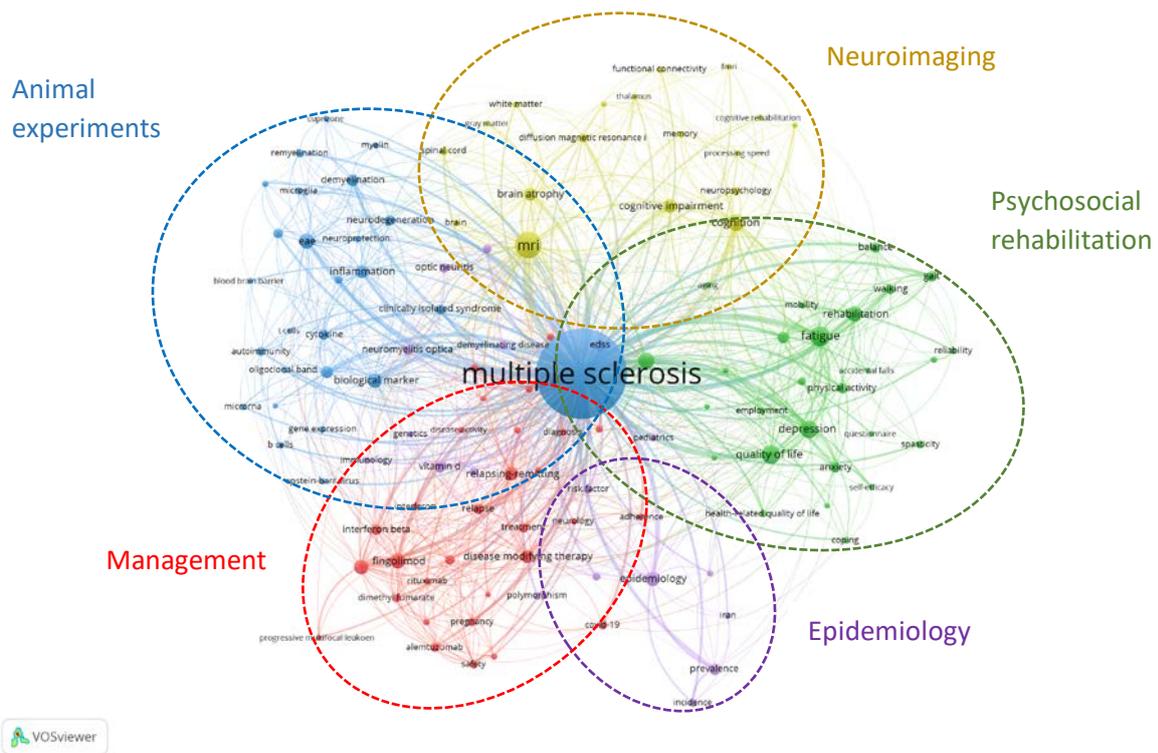

**Figure 3.** The main clusters of MS studies based on keyword co-occurrence analysis.

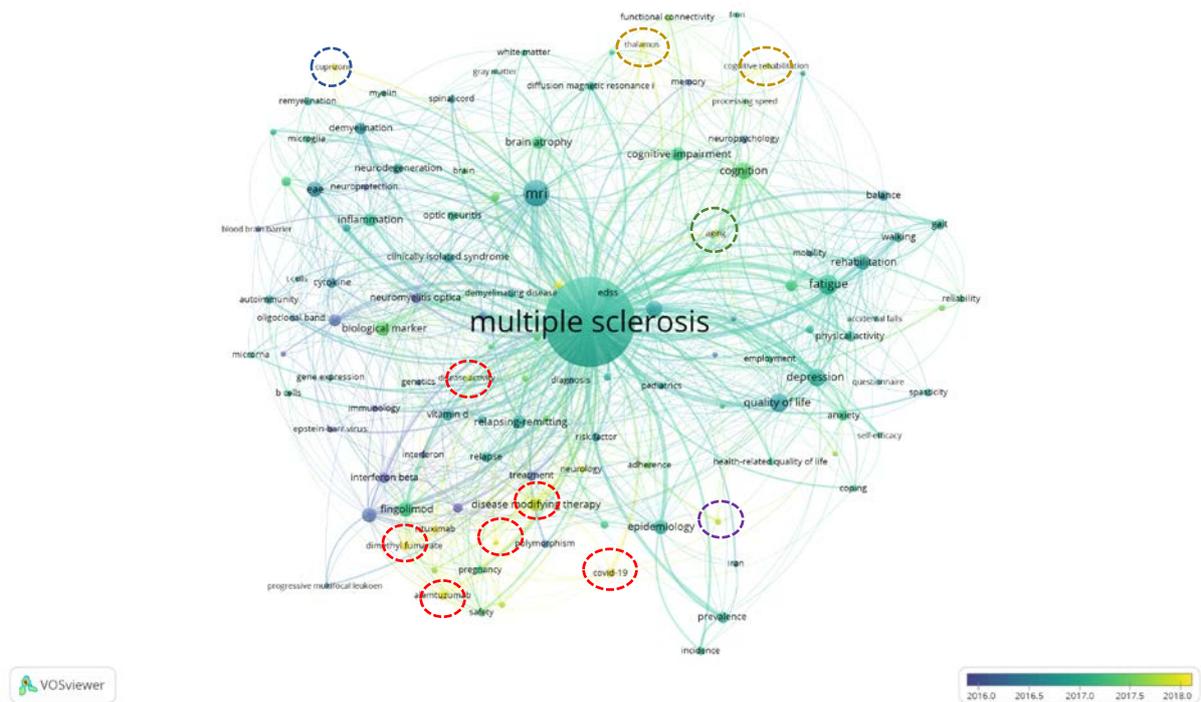

**Figure 4.** Temporal overlay on keyword co-occurrence map.

The ESDs were then generated by comparing the total link strength (centrality), occurrence (density), and average year of publication (novelty) of each keyword to their derived median values. Based on the plane's position, four types of themes can be determined.[11,12] Figure 5(a) depicts the four themes in the novel publication year. These are: emerging with high density (upper-left quadrant), emerging with low density (lower-left quadrant), core (upper-right quadrant), and interdisciplinary (lower-right quadrant). Figure 5(b) shows the four themes in the old publication year, consisting of isolated (upper-left quadrant), obsolete (lower-left quadrant), mature (upper-right quadrant), and declining (lower-right quadrant).

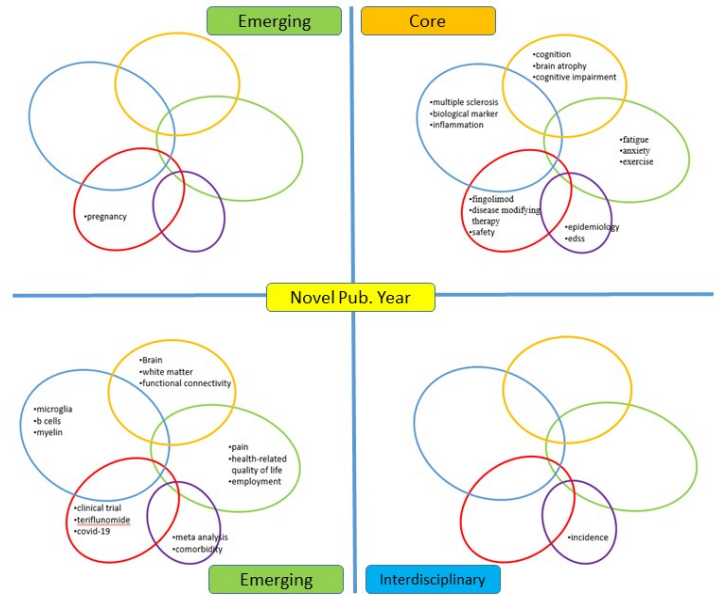

a

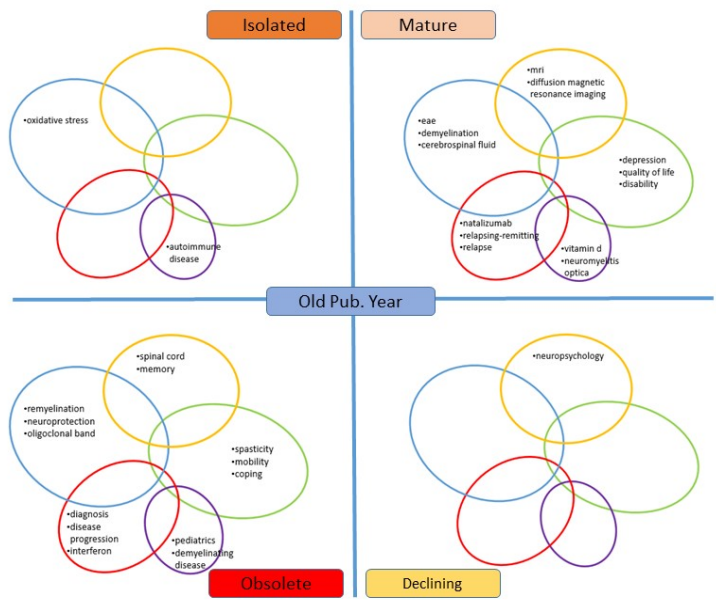

b

**Figure 5.** Enhanced strategic diagrams showing (a) emerging, core and interdisciplinary in the novel publication year (b) isolated, mature, obsolete and declining topics in the old publication year.

**Discussion**

This bibliometric analysis found 18003 Scopus-indexed articles published between 2012 and 2021. According to our data, research output increased gradually from 2012 to 2019, then exploded in 2019 and continued to the current year. When compared with the earlier bibliometric analysis on MS,[6] the number of articles has almost doubled. Only 9778 articles were published between 2003 and 2012,[6] while in almost the same duration, 18003 articles were published between 2012 and 2021 and this number is expected to increase by end of 2021. However, the overall growth rate of publications was similar i.e. around 82%.

Science mapping capability to portray the conceptual structure of scientific subjects is unique.[16] In this study, we found five conceptual clusters (themes) in MS literature: management, psychosocial rehabilitation, animal experiment, neuroimaging, and epidemiology. Cluster 1 (management) has the most terms (28), followed by Cluster 3 (animal experimentation, 22), Cluster 2 (neuropsychological rehabilitation, 19), Cluster 5 (epidemiology, 18) and Cluster 4 (neuroimaging, 11 keywords). This could indicate that MS management research is currently the most popular and has the biggest volume.[17]

Our results identified several core topics in each cluster/theme of MS research. These topics were the most frequent and had strong external links. For example, "fingolimod", disease-modifying therapies" and "safety" were the core topics under management theme. Many studies related to these topics were published recently.[18-20] Under animal experiment theme, "multiple sclerosis", "biological marker", and "inflammation" were the core topics. Biological markers were useful in the context of inflammatory disorders to precisely describe the immune response and prospective therapeutic targets, as well as to better understand the etiopathogenesis and monitoring of disease activity and treatment response.[21] They play a diagnostic role[22] as well as predict disability progression, monitor ongoing disease activity and assess treatment response in MS.[23] Under the neuropsychological rehabilitation theme, the core keywords were "fatigue", "anxiety" and "exercise". While new disease-modifying medications are useful in slowing disease progression and neurological decline, there is no effective pharmacological treatment for fatigue.[24] Physical exercise significantly reduces fatigue[26] in MS patients, but there is insufficient data support for anxiety.[26]

The core keywords under neuroimaging theme were "cognition", "cognitive impairment", and "brain atrophy". Cognitive impairment is present in all MS phenotypes,[27] with a prevalence of 43–70% depending on phenotype and cognitive diagnostic criteria utilised.[28] Structural correlates of cognitive impairment include white matter and grey matter damage, brain atrophy and network modifications.[29] Under the epidemiology theme, the core keywords were "epidemiology", and "edss". Expanded Disability Status Scale (EDSS) is a commonly used measure of disability for MS in various studies including epidemiology studies. [30,31]

The emerging keywords in MS literature include "covid-19", "teriflunomide", "clinical trial", "microglia", "b cells", "myelin", "brain", "white matter", "functional connectivity", "pain", "employment", "health-related quality of life", "meta-analysis" and "comorbidity", while "pregnancy" was the emerging keyword with higher density. These keywords has the potential to expand. For example, coronavirus disease-19 (COVID-19) has become a major concern for medical professionals across the board including management of MS patients who are often on immunosuppressive drugs. With the pandemic, the therapeutic landscape for MS has shifted, and it will undoubtedly continue to do so in the near future.[32] While clinical trial and teriflunomide were the emerging topics under management theme, based on the case reports, teriflunomide was safe and used as the first-line treatment for MS patients with COVID-19.[33]

The emerging keywords in the animal experiment theme include "B cells", that has recently been identified as the main mechanism in the pathogenesis of MS,[34] and "microglial", which has a role in rescuing oxidized phosphatidylcholines (oxPC)-induced neurotoxicity[35] in MS. Under the psychosocial rehabilitation theme, the emerging keywords were "health-related quality of life", "pain", and "employment". Various symptoms including pain affect the employment status and quality of life of working-age adults with MS.[36] Furthermore, according to a recent study, the impact of MS on health-related quality of life (HRQOL) in real-world patients may be underestimated.[37] A new review also emphasises the need for rehabilitation in fostering daily life involvement in MS patients, as part of a comprehensive plan to address present and future issues.[38]

"Brain", "white matter" and "functional connectivity" were the emerging keywords under neuroimaging theme. Several diffusion magnetic resonance imaging (MRI) studies have found that MS patients have lower structural connectome,[38] which is best explained by the disruption of long-range white matter tracts[39] and the disconnection of the major hubs such as the thalamus and nodes of the default-mode brain network.[40] A functional connection such as resting-state functional MRI (rs-fMRI) was another growing area under this theme and is increasingly being employed to investigate the aetiology of cognitive impairment in MS patients.[41]

Finally, under the epidemiology theme, the most frequent keywords were "meta-analysis" and "comorbidity". In recent years, comorbidities in patients with MS have piqued researchers' interest.[42] Meta-analysis was frequently conducted to pooled epidemiological data such as worldwide prevalence of MS in specific regions, for example, in Asia and Oceania, [43] or in paediatric age group,[44] as well as in other related issues such as suicidal ideation.[45]

In conclusion, this study has demonstrated a rapid expansion of worldwide literature on MS, with the development trend predicting that it will more than double in size during the next decade. Based on the topic analysis using ESDs, we proposed several emerging research topics under each identified theme, namely management, psychosocial rehabilitation, animal experiment, neuroimaging, and epidemiology, which will likely expand in the near future.

**Declaration of Conflicting Interests**

The Authors declare that there is no conflict of interest.

**Funding**

The authors would like to acknowledge the Ministry of Higher Education Malaysia for Fundamental Research Grant Scheme with Project Code: FRGS/1/2020/SKK0/USM/02/29 for the financial support.

**ORCID iDs**

| | |
|---|---|
| Nazlahshaniza Shafin | https://orcid.org/0000-0002-8524-1852 |
| Che Aishah Nazariah Binti Ismail | https://orcid.org/0000-0003-1723-3575 |
| Mohd Zulkifli Bin Mustafa | https://orcid.org/0000-0002-8140-8184 |
| Nurhafizah Binti Ghani | https://orcid.org/0000-0002-5596-7548 |
| Asma Hayati Ahmad | https://orcid.org/0000-0001-5447-0356 |
| Zahiruddin Othman | https://orcid.org/0000-0002-9070-2078 |
| Adi Wijaya | https://orcid.org/0000-0001-5339-0231 |

Rahimah Zakaria  https://orcid.org/0000-0002-2459-321